\documentstyle[12pt,epsf]{article}
\newcommand{\ds }{\displaystyle}
\newcommand{\ra}{\rightarrow}
\newcommand{\be}{\begin{equation}}
\newcommand{\ee}{\end{equation}}
\newcommand{\bea}{\begin{eqnarray}}
\newcommand{\eea}{\end{eqnarray}}
\newcommand{\ci}{\cite}
\newcommand{\bi}{\bibitem}
\newcommand{\nono}{\nonumber \\}

\newcommand{\dd}{\partial}

\newcommand{{\bfna}}{\mbox{\boldmath$\vec{\nabla}$}}

\newcommand{\k}{\tilde{k}}
\newcommand{\ti}{\tilde{t}}
\newcommand{\m}{\tilde{\mu}}

\def\dal{\,\lower0.3ex\vbox{\hrule\hbox{\vrule\kern2pt\vbox{\kern4pt\kern4pt}
\kern2pt\vrule}\hrule}\,}

\def\o{\omega}

\begin{document}

\title{\sl Assisted Tunneling of a metastable state between barriers}
\vspace{1 true cm}
\author{G. K\"albermann$^*$
\\Soil and Water dept., Faculty of
Agriculture, Rehovot 76100, Israel}
\maketitle

\vspace{3 true cm}
\begin{abstract}

The assisted tunneling of a metastable state between barriers is investigated 
analytically by means of a simplified one dimensional model. 
A time dependent perturbation changes the pole spectrum
of the wave function introducing a larger decay constant.
New insights about the decay of a metastable state are found.
The scheme is exemplified for parameters
corresponding to the nuclear process of $\alpha$ decay.
\end{abstract}

$^*${\sl e-mail address: hope@vms.huji.ac.il,germankal@hotmail.com}

\newpage
\section{\sl Introduction}
Quantum tunneling is the conventional theoretical paradigm for the explanation
of a gamut of atomic, molecular and nuclear phenomena.
Classically forbidden regions can be accessed by a quantum object whose
behavior is described in terms of a wave function.

Shortly after the advent of quantum mechanics, Gurney and Condon and 
simultaneously Gamow, explained the huge differences between 
$\alpha$ decay lifetimes of similar unstable nuclei using the concept of 
tunneling through a barrier.\ci{condon, gamow}
This framework has remained as reliable today as when it was proposed
almost a century ago. \ci{medeiros, holstein}
The question of assisting tunneling, i.e. accelerating or decelerating
the decay process is of the utmost importance in the nuclear context and 
elsewhere. In the nuclear case, acceleration of the decay process could
change dramatically the treatment of radioactive waste as well as providing
alternative fuel sources.

Analytical expressions for the various tunneling processes
are extremely important. It is virtually impossible to follow numerically the 
evolution of a wave function to the long times involved in decay processes.
There is a huge gap between the natural time scale of nuclear phenomena of 
the order of $10^{-24} sec$ and the decay time scale of the order of 
milliseconds to millions of year.\ci{van dijk} 
The situation is somewhat, but not radically
different, for atomic phenomena. 

Not surprisingly, 
assisted tunneling is extensively investigated for solid state systems.
The main theoretical tool in this endeavor is the
Floquet formalism\ci{shirley}. 
This method involves large matrices and it is
computationally quite intensive. \ci{hanggi}.   
Other methods such as the "elevator effect" model 
of resonance assisted activation attempt to attack the problem in
a semi-analytic manner\ci{azbel}.
Ivlev\ci{ivlev} has recently developed a complex time method to investigate
assisted tunneling, aimed at a nonperturbative approximate 
treatment of the tunneling through special barriers.

There seems to be a clear consensus, that exposing a system to external
excitations, time harmonic or not, can enhance its tunneling rate.
The absence of exact analytical results, and reliable long term
numerical results, especially for the nuclear case, makes difficult 
the assessment of experimental feasibility of implementing assisted activation.
 
In the present work we investigate 
an extremely simple model of a one dimensional
quantum mechanical system located initially between fixed barriers.
A time harmonic potential such as a low frequency -as compared to
the natural frequency of the system-  electric field, is then
applied. The simplicity of the model allows a full, albeit perturbative,
exact analytical solution. 
The perturbation causes a qualitative change 
in the pole structure of the wave function. The spectrum of poles arising from
the normalization of the wave function that rule the decaying behavior of
the metastable state, is extended.
These poles in the complex momentum space, appear as the wave function
is expanded in a complete set of the unperturbed system.
The poles contribute in a contour integration of the wave amplitudes
in momentum space.\ci{holstein}. They are not introduced ad-hoc, 
as Gamow did with his complex energy method, but, arise in a systematic
expansion of the wave function naturally.
The time development of the wave function is dominated by 
the lowest energy poles.
The introduction of the perturbation opens a new line of poles. Here again
only one of them appears important. This new pole produces a bigger decay 
constant. 

The result is not specific to time varying fields. 
The new line of poles arises even for a static spatially inhomogeneous
potential such as that of a constant electric field.
In a nuclear setting, such a static field will be certainly almost
completely screened by the atomic electrons, whereas for 
a time varying potential, the efficiency of such screening is diminished. 
A discussion of the electronic screening 
problem is beyond the scope of the present work.
We henceforth treat only a time harmonic perturbation that vanishes in the
zero frequency limit. 

The external vibration shakes the system back and 
forth between barriers and accelerates the tunneling process.
We analyze the general and, mostly qualitative results, based on the pole
spectrum. New insights into some troubling aspects of metastable state 
decay are offered in section 5.
Extensive mathematical details will be presented in a forthcoming work.

\section{\sl The Model}
The Schr\"odinger equation for the one dimensional system
located in between barriers is taken to be

\be\label{h1}
i~\frac{\dd\Psi}{\dd t}=\frac{-1}{2~m}\frac{\dd^2 \Psi}{\dd x^2}+
\lambda~(\delta(x+x_0)+\delta(x-x_0))\Psi
\ee

\noindent Comparing to a finite size square barrier, 
$\ds \lambda$ represents the height multiplied by the width. 
For nuclear $\ds \alpha$ decay $\ds \lambda\approx~2$,
$\ds ~m\approx~20 fm^{-1}$,$\ds ~x_0\approx~10~fm, m\lambda~x_0=~400 >> 1$

We choose as initial nonstationary wave, a gaussian wave packet. 
This allows a full analytical treatment. In $\ds \alpha $ decay, it
would correspond to a preexisting spatially symmetric state.

\bea\label{packet}
\Psi(x,t=0)=N~e^{-\frac{x^2}{\Delta^2}}
\eea

\noindent with N a  constant factor normalization, 
and $\ds \Delta$ is the width parameter of the packet.
Inside the barrier region the packet disperses if $\ds \Delta~<~x_0$.
, but, soon enough it stops
dispersing due to the presence of the barriers. It can then
spread only through the tunneling process governed by the barriers.

The even ({\sl e}) and odd ({\sl o}) stationary states of eq.(\ref{h1}) 
are readily found to be

\bea\label{even}
n_e(k)~\chi_e(x)=\left \{
\begin {array}{rl}
cos(k~x)&if~|x|<x_0\\
A~cos(k~x)\pm B sin(k~x)& if~ x>x_0~or~x<-x_0
\end{array}\right.
\eea

\bea\label{odd}
n_o(k)~\chi_o(x)=\left \{
\begin {array}{rl}
sin(k~x)& if~|x|<x_0\\
\pm C~cos(k~x)+ D sin(k~x)&if~x>x_0~or~x<-x_0
\end{array} \right.
\eea

The set of even-odd functions is orthonormal and complete. The
normalization factors\footnote{The determination of the 
normalization and the completeness issue 
will be dealt with in an extended version of the paper.\ci{trott}}
are extremely important and determine the location of the poles in the complex
plane. As shown below, these poles govern the exponential decay of 
metastable wave functions. 

The  normalization factors are

\bea\label{neven}
(n_e(k))^2&=&\pi~(A(k)^2+B(k)^2)\nono
A(k)&=&1-sin(2~k~x_0)~\frac{m\lambda}{2~k} \nono
B(k)&=& \frac{m\lambda}{k}~(cos(k~x_0))^2\nono
k^2~(n_e(k))^2&=&\pi~\bigg((k-sin(2~k~x_0)~\frac{m\lambda}{2})^2+
(m\lambda~(cos(k~x_0))^2)^2\bigg)
\eea

\bea\label{nodd}
(n_o(k))^2&=&\pi~(C(k)^2+D(k)^2)\nono
C(k)&=&-\frac{m\lambda}{k}~(sin(k~x_0))^2\nono
D(k)&=&1+sin(2~k~x_0)~\frac{m\lambda}{2~k}\nono
k^2~(n_o(k))^2&=&\pi~\bigg((k+sin(2~k~x_0)~\frac{m\lambda}{2})^2+
(m\lambda~(sin(k~x_0))^2)^2\bigg)
\eea

The time harmonic potential perturbation reads

\bea\label{potential}
V(x,t)=\mu~x~sin(\o~t)
\eea

\noindent with $\ds \mu$ a coupling constant. For an external
electric field of intensity $\ds E_0$ interacting with an $\ds \alpha$ 
particle of charge $\ds 2~|q_e|,~\mu=2~|q_e|~E_0$.

 The wave function is expanded in the complete set of even and odd states

\bea\label{expansion}
\Psi(x,t)=\sum_{i=e,o}\int_0^{\infty}\chi_i(k,x)~a_i(k,t)~e^{
\frac{-i~k^2~t}{2~m}}~dk
\eea
\noindent dots denoting derivatives with respect to {\sl t}.

The Schr\"odinger equation for the amplitudes $\ds a_{e,o}$ becomes

\bea\label{sch1}
i~\dot{a}_e(k)&=&\int{e^{-\frac{(k'^2-k^2)~t}{2m}}
<\chi_e(k,x)|~V(x,t)|\chi_o(k',x)> dk'~a_o(k',t)}\nono 
i~\dot{a}_o(k)&=&\int{e^{-\frac{(k'^2-k^2)~t}{2m}}
<\chi_o(k,x)|~V(x,t)|\chi_e(k',x)> dk'~a_e(k',t)}
\eea

The leading contribution to the matrix element of the interaction can
be evaluated exactly
 
\bea\label{melement}
<\chi_e(k,x)|~V(x,t)|\chi_o(k',x)>&=&\int_{-\infty}^{\infty}
{\chi_e(k,x)~V(x,t)\chi_o(k',x)~dx}\nono
&\approx&\mu~\pi\frac{1}{n_e(k)~n_o(k')}
\frac{\dd\delta(k-k')}{\dd k'}~sin(\o~t)
\eea

Inserting eq.(\ref{melement}) in eq.(\ref{sch1}) we obtain

\bea\label{sch2}
i~\dot{a}_e(k,t)=-\frac{\pi}{n_o(k)~n_e(k)}
(a_o'(k,t)-n_o'(k)/n_o(k)~a_o(k,t)-\frac{i~k~t}{m}a_o(k,t))\mu~sin(\o~t)\nono
i~\dot{a}_o(k,t)=-\frac{\pi}{n_o(k)~n_e(k)}
(a_e'(k,t)-n_e'(k)/n_e(k)~a_e(k,t)-\frac{i~k~t}{m}a_e(k,t))\mu~sin(\o~t)
\eea
\noindent primes denoting derivatives with respect to {\sl k}.

Rescaling to dimensionless variables

\bea\label{rescal}
k&\ra&\k=\frac{k}{\sqrt{m~\o}}\nono
t&\ra&\ti=\o~t\nono
\mu&\ra&\m=\frac{\mu}{\sqrt{\o^3~m}}
\eea

\noindent eq.(\ref{sch2}) becomes

\bea\label{sch3}
i~\dot{a}_e(\k,\ti)=-\m\frac{\pi}{n_o(k)~n_e(\k)}
(a_o'(\k,\ti)-n_o'(\k)/n_o(\k)~a_o(\k,\ti)-{i~\k~\ti}~a_o(\k,\ti))
~sin(\ti)\nono
i~\dot{a}_o(\k,\ti)=-\m\frac{\pi}{n_o(\k)~n_e(\k)}
(a_e'(\k,\ti)-n_e'(\k)/n_e(\k)~a_e(\k,\ti)-{i~\k~\ti}~a_e(\k,\ti))
~sin(\ti)
\eea

\noindent primes denoting derivatives with respect to $\ds \k$ and dots
representing derivatives with respect to $\ds \ti$.

$\m$ is the relevant parameter of the problem. It 
is small for external angular frequencies
$\ds \o_{min}^3> \mu^2~m$. \footnote{
For example, in $\ds \alpha $ decay case with $\mu\approx
1~eV^2$ corresponding to a strong electric field amplitude of 
$\ds E_0\approx 5~10^6~Volt/m$, $\ds \o_{min}>10^{12}~sec^{-1}$.}
For lower frequencies, nonperturbative solutions are required.
The smaller the electric field, the lower the frequency 
for which the perturbative solution will be appropriate.

The amplitudes of eq.(\ref{sch3}) are further expanded in powers of $\ds \m$

\bea\label{ampl}
a_e(\k,\ti)=\sum_{n=0}^{\infty}~b_{e}^{(n)}(\k,\ti)~\m^n\nono
a_o(\k,\ti)=\sum_{n=1}^{\infty}~b_{o}^{(n)}(\k,\ti)~\m^n
\eea

\noindent with initial conditions

\bea\label{conditions}
b_{e,o}^{(n)}(\k,t=0)&=& 0~for~n\ge 1\nono
b_e^{(0)}(\k,\ti) &=& b_0(\k)/n_e(\k)=\int_{-\infty}^{\infty}
N~e^{-\frac{x^2}{\Delta^2}}\chi_e(k) dx
\eea

\noindent We extracted the normalization factor of the even
wavefunctions $\ds n_e(k)$ for the sake of convenience.
It is straightforward to show that all the odd {\sl n} powers of $\ds 
a_e(\k,\ti)$ vanish identically, and the same is true 
for the even {\sl n} powers $\ds a_o(\k,\ti)$.
The substitution of eqs.(\ref{ampl}) into eq.(\ref{sch3}) 
produces a separate equation for each 
order {\sl n}. The equations are integrable exactly order by order, although
the expressions of higher order amplitudes become increasingly involved.

\section{\sl Poles of the Wave Function}
Figure 1 shows $\ds \frac{\pi}{k^2~n_e(k)^2}$, and
$\ds \frac{\pi}{k^2~n_p(k)^2}$ , for the parameters 
$\ds m\lambda~x_0=400, x_0=10 fm$.
The thin and tall spikes are due to the extreme closeness of the minima of
the normalization factors to their complex zeroes.

\begin{figure}
\epsffile{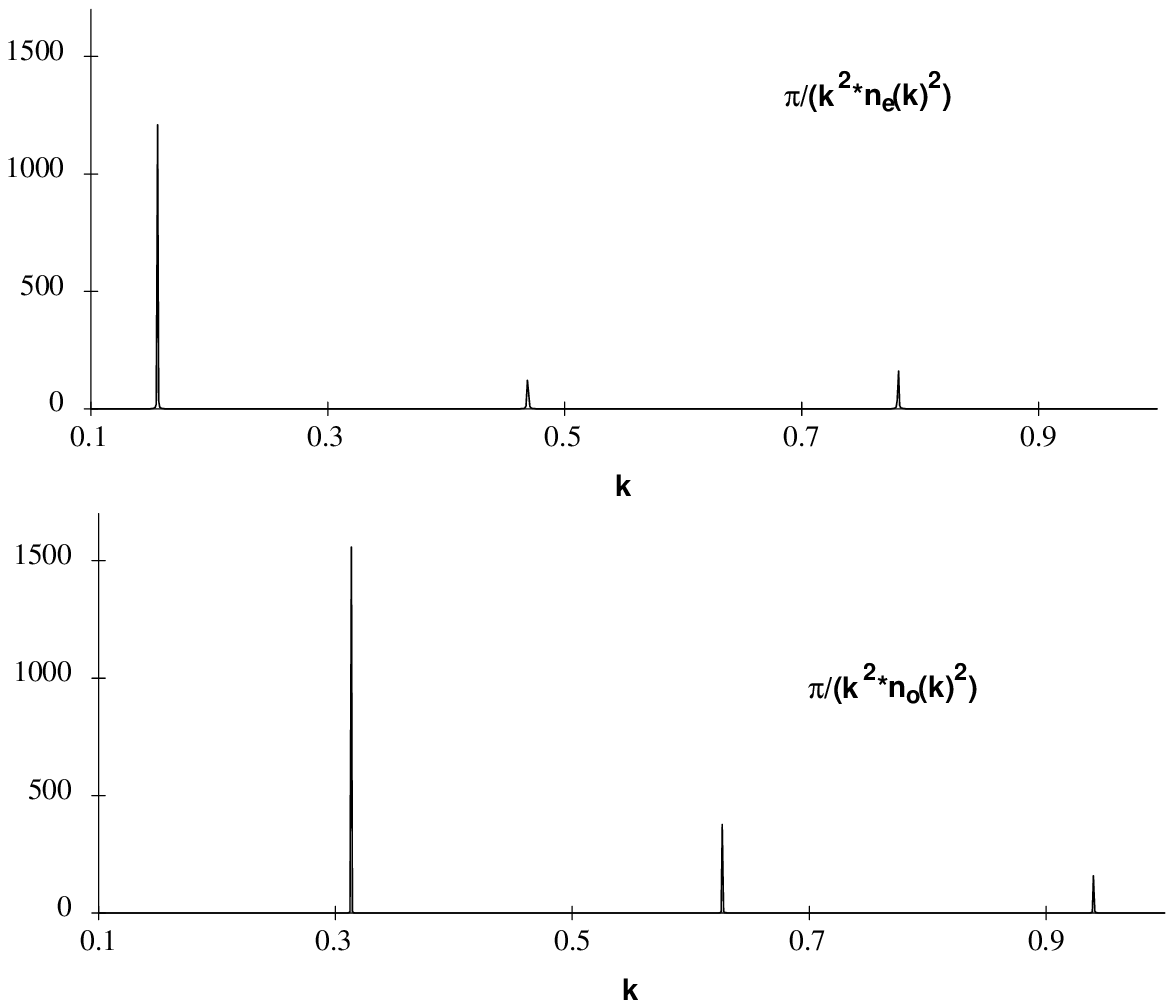}
\caption{$ \frac{\pi}{k^2~n_e(k)^2}$, and
$\frac{\pi}{k^2~n_p(k)^2}$ , as a function of {\sl k} in units of $fm^{-1}$ 
for the parameters $m\lambda~x_0=400, x_0=10 fm$.}
\label{fig1}
\end{figure}

The amplitudes of eq.(\ref{ampl}) determine the wave function 
through eq.(\ref{expansion}).
Integration over {\sl k} is dominated by the pole structure of the
normalization factors $\ds n_{o,e}(k)$.

The normalization factors of eq.(\ref{neven},\ref{nodd}) 
are even in the argument {\sl k}.
Consequently, the behavior around their zeroes 
can be expressed in the form

\bea\label{pole1}
k^2~n_{e,o}^2&\approx&~\gamma_{e,o}^{(j)}~(k^2-{k^{(j)}_{e,o}}^2)^2
+\beta_{e,o}^{(j)}
\eea
\noindent where {\sl j} enumerates the pole number.

For the $\delta$ barriers we have chosen, and in the limit of 
$m\lambda~x_0~>>~1$, we can find simple
expressions for the locations of the poles as well as the value of
the norm factors on the real  axis at $\ds k = k^{(j)}_{e,o}$. 

\bea\label{beta}
k_e=\frac{(2n+1)~\pi~m\lambda}{2~(1+m\lambda~x_0)}~&~&~
k_o=\frac{n~\pi~m\lambda}{(1+m\lambda~x_0)}\nono
\gamma_e=\pi\frac{2~m\lambda~x_0^3(m\lambda~x_0+4)}{(2n+1)^2\pi^2}~&~&~
\gamma_o=\pi\frac{m\lambda~x_0^3(m\lambda~x_0+4)}{4~n^2\pi^2}\nono
\beta_e=\frac{(2~n+1)^4~\pi^4}{16~m^2\lambda^2~x_0^4}~&~&~
\beta_o=\frac{n^4~\pi^4}{m^2\lambda^2~x_0^4}
\eea

The  poles $\ds q_n,q_m$ are located symmetrically above and below the real 
momentum axis at $k~x_0$ approximately equal to an odd multiple of
$\ds \frac{\pi}{2}$ for the even case and, and a multiple of $\ds \pi$ for the odd 
case, where the bound states of the infinite wall case lay.

\bea\label{poles}
q_n^e&=&\bigg(\frac{(2n+1)\pi~ m\lambda}{2~(1+m\lambda~x_0)}\bigg)^2
\pm~i\sqrt{\frac{\beta_e}{\gamma_e}}\nono
q_o^2&=&\bigg(\frac{n\pi~ m\lambda}{(1+m\lambda~x_0)}\bigg)^2\pm i~
\sqrt{\frac{\beta_o}{\gamma_o}}
\eea

The imaginary parts of the poles are orders of magnitude smaller than the
real parts. For example, inserting the parameters corresponding to
$\ds \alpha$ decay, the first even pole appears at $\ds q_{e,n=1}\approx 0.024
\pm i~3 * 10^{-7} fm^{-2}.$ For a finite size width, the imaginary parts are 
much smaller.
The sharp spikes seen 
in figure 1 lead to the dominance of the poles in the spectrum of momenta.

When eq.(\ref{expansion}) is evaluated by contour integration
in the complex $\ds k^2$ plane, the contour has to be closed
from below. In the lower half-plane the convergence is insured by 
the exponential $\ds e^{-i~\frac{k^2~t}{2~m}}.$

The poles are separated from each other, their contribution adds up.
The negative imaginary part of each pole induces a time decaying exponential.
Each exponent determines a different decay constant and decay time.
If the original wave function is even in space, as in the expression of 
eq.(\ref{packet}),only {\sl even} poles contribute to the unperturbed
decay process.
Due to the initial confinement of the wave to the inter-barrier region,
the most important contribution arises from the first even pole.
The influence of higher order poles is hindered by the wave packet transform to
momentum space 

\bea\label{hind}
\Psi(k)\sim e^{-\frac{k^2~\Delta^2}{4}}
\eea
 
Therefore, the first pole drives 
the decay process and determines the lifetime of the quasistationary state.

\section{\sl Acceleration of the Decay Process}

The structure of the solution of eq.(\ref{sch2}) 
for the amplitude series of eq.(\ref{ampl}) and to the lowest order 
in $\ds \mu$ for the even amplitude is found to be 

\bea\label{aeven}
a_e(\k,\ti)&=&n_e(\k)~b_0(\k)\bigg(~\frac{1}{n_e^2}
+\m^2~\tilde{b}_e^{(2)}(\k,\ti)\bigg)\nono
\tilde{b}_e^{(2)}(\k,\ti)&=&\frac{e_1(\ti)+e_2(\ti)~\k^2}{n_e^4~n_o^2}
+\frac{e_3(\ti)~\k~n'_o}{n_e^4~n_o^3}\nono
&+&\frac{\k~e_4(\ti)~n_e'+e_5(\ti)~n_e''}{n_e^5~n_o^2}\nono
&+&\frac{e_6(\ti)}{n_e^6~n_o^2}+\frac{e_7(\ti)n_e'n_o'}{n_e^5~n_o^3}
\eea

\noindent The odd amplitude reads

\bea\label{aodd}
a_o(\k,\ti)&=&n_o(\k)~b_0(\k)~\m~\tilde{b}_o^{(1)}(\k,t)\nono
\tilde{b}_o^{(1)}(\k,\ti)&=&\frac{k~o_1(\ti)}{n_e^2~n_o^2}
+\frac{o_2(\ti)~n'_e}{n_e^3~n_o^2}
\eea

\noindent where we have extracted a normalization factor in advance
of the integration over the even-odd set of wave functions $\chi(k)$, as we want
to identify the influence of the pole structure on the amplitudes. In 
eqs.(\ref{aeven},\ref{aodd}) 
we have suppressed the $ \k$ dependence of the normalization factors in
the denominators for brevity.

We can readily analyze the structure of the perturbed amplitudes.
The amplitudes in eqs.(\ref{aodd},\ref{aeven}) differ
from the unperturbed amplitude $\ds b_0(k)$ by the appearance of poles
for the odd set of wave functions introduced by powers
of $n_o(k)$ in the denominators.

The first even pole of eq.(\ref{beta}) generates a decay constant
of the form
\bea\label{dece}
|e^{-i\frac{k_{e,n=1}^2~t}{2~m}}|&\ra&~e^{-\Lambda_e~t}\nono
\Lambda_e&=&\frac{1}{2~m}\sqrt{\frac{\beta_1}{\gamma_1}}\nono
&=&\frac{\pi^3}{8~m~x_0^4~m^2~\lambda^2}
\eea

\noindent whereas the first odd pole generates a decay constant 

\bea\label{deco}
\Lambda_o=8~\Lambda_e
\eea
\noindent The factor of {\sl 8} is specific to the $\delta$ model. In general
there should be roughly this order of magnitude increase in the decay constant.

At the same time, the real part of the pole
that enters the wave function mainly through the momentum space transform of
the initial wave function of eq.(\ref{hind}) changes also

\bea\label{damp}
e^{-\frac{\pi^2~\Delta^2}{16~x_0^2}}&\ra&
e^{-\frac{\pi^2~\Delta^2}{4~x_0^2}}\nono
e^{-\frac{\pi^2~\Delta^2}{16~x_0^2}}&\ra&e^{-\frac{9\pi^2~\Delta^2}{16~x_0^2}}
\eea

\noindent The left hand side in both lines in eq.(\ref{damp})
correspond to the first even pole contribution
whereas the right hand sides pertain to the first odd pole and the 
second even pole respectively.

There arise then two effects: An enhancement of the decay constant by a factor
of {\sl 8} and, a damping factor.
A rough estimate of the importance of the news terms can now be performed.

If initially $\Delta\approx\ds x_0$, then the first odd pole contributes  
to the wave function about $10\%$, while the second even pole
contributes much less than $\ds 1\%$. Even without getting into the 
cumbersome details of
the wave function, the survival probability in the inner region between
barriers will be influenced markedly by the introduction of the
perturbation. The wave will tunnel faster.

\section{\sl Flux Continuity}

The question of conservation of probability or flux continuity seems
unavoidable at this junction. This problem arises even before adding
a harmonic perturbation.
If the wave diminishes with time everywhere it appears as if
unitarity is broken. This should not occur in the evolution of 
a wave function with the Schr\"odinger equation and a real potential.
The contradiction arises in other conventional approaches 
to the decay of a metastable state\ci{holstein}. In the scattering method
the paradox is apparently resolved by exhibiting 
another piece in the wave function that increases
exponentially with time. 

However, in the present treatment of the problem, the wave function 
receives contributions
from the poles in the lower momentum half-plane only. There seems to be no
time increasing piece at all. The exponentials originating from those
poles is decreasing in time.
On the other hand, flux has to be continuous 
and overall probability should be conserved.
The answer to this conundrum brings to
the fore the mathematical beauty of the decay process.

Consider the two separate regions, the inside zone around $x\approx~0$ and the
long distance zone of $\ds x\ra\infty$, far away from the microscopic
location of the barriers.
In the inner region the integration over momentum space is dominated
essentially by the poles. This is true until very long times. 
At very
long times the exponential factor $\ds e^{-i\frac{k^2~t}{2 m}}$
oscillates wildly for small changes in the momentum. The integral
of eq.(\ref{expansion}) will start to suppress the importance of the pole 
share in the integral. The pole spike will be twisted by the oscillating
factor with alternating positive and negative parts.
It is possible to estimate the time at which the assumption of pole dominance
fails as follows.
The width of the pole spikes read off
from eq.(\ref{pole1}) is $\ds \delta(k^2)\approx\gamma$. For $\ds
k^2\approx\gamma$ the inverse of the normalization factor essentially
vanishes  $\ds \frac{1}{n_{o,e}^2}\ra 0$.
The pole dominance will end when

\bea\label{collapse}
\frac{k^2~t}{2~m}\ra\frac{\gamma~t}{2~m}\approx 1
\eea

\noindent Inspection of the values of $\ds \gamma$ listed in eq.(\ref{beta}),
it is found that this time is of the same order of the 
lifetime $\ds \frac{1}{\Lambda}$.
Hence, in the inner region, the wave will decay until a time of the order
of the lifetime. After that time, other contributions to the wave
arising from momenta outside the poles have to be included. 
The wave leaks out of the inner region 
for very long times of the order of the lifetime of the state.

The behavior in the long distance region is completely different, even for
short times.
For $\ds x\ra\infty$ the oscillations in the integrand of 
eq.(\ref{expansion}) stem from the harmonic functions {\sl in space} $\ds\chi$
of eqs.(\ref{even},\ref{odd}).
For the long distance region, 
{\sl the poles will never dominate} , neither at short times
nor at asymptotically long times.
The wave is built from the whole momentum spectrum. 
Moreover,
as time passes and, in the spirit of the stationary phase approximation, 
smaller and smaller momenta will dominate.
This effect will {\sl enhance} the wave at long distances, due to the 
momentum space factors of eq.(\ref{hind}).
Consequently, the wave function will decay in the inner region and grow 
in the outer region and probability will be conserved.
The same is true when more than one decay constant is present.

\section{\sl Conclusions}
The investigation of assisted tunneling carried out
in the present work implies that, there are distinctive qualitative 
features induced by the introduction of a time dependent perturbation.
The main signal for the process is the appearance
of another decay constant, bigger than the one of the unperturbed case. 
The state decays as if it were composed of two
channels with different partial decay widths. 
A large component
with a longer lifetime and a small, but presumably non-negligible component,
with a much smaller -eight times smaller for $\ds m\lambda~x_0~>>~1$-
lifetime.
Evidently, the fine details are important for an accurate
determination of the wave function, and will be shown in a coming
publication.
The reverse process of tunneling into a region enclosed between barriers
could be affected by the mechanism of assisted tunneling also. 
Such a possibility is of relevance
for atomic and nuclear processes, such as nuclear fusion.

\end{document}